\documentclass[amsmat,amssymb,amsfonts,aps,prb,twocolumn,showpacs]{revtex4}

\usepackage{graphicx}
\usepackage{dcolumn}
\usepackage{bm}
\begin{document}

\title{Electronic Structure of gated graphene and graphene ribbons}

\author{J. Fern\'andez-Rossier$^{1}$
, J. J. Palacios$^{1}$ ,and L.Brey$^{2}$. }
\affiliation{(1)Departamento de F\'{\i}sica Aplicada, Universidad de
Alicante, Spain \\
(2)Instituto de Ciencia de Materiales de Madrid, CSIC, Spain  }

\date{\today}

\begin{abstract}
We study the  electronic structure of  gated  graphene sheets. We
consider both infinite graphene and finite width ribbons. The effect
of Coulomb interactions between the electrically injected carriers and
the coupling to the external gate are computed self-consistently in
the Hartree approximation.  We compute the average density of extra
carriers, $n_{2D}$, the number of occupied subbands  and the density
profiles as a function of the gate potential $V_g$. We discuss
quantum corrections to the classical capacitance and  we calculate
the threshold $V_g$ above which   semiconducting armchair ribbons
conduct. We find that the ideal conductance of perfectly
transmitting wide ribbons is proportional to the square root 
of the  gate voltage.
\end{abstract}

\maketitle

\section{Introduction}
One important ingredient of semiconductor microelectronics  is  the
electrical tunability of the resistance through the  field effect.
The recent demonstration of field effect in
graphene\cite{Geim05,Kim05,Science06} has opened  a new research
venue. Graphene is different from conventional semiconductors for a
number of reasons.  First, graphene is a truly two dimensional
atomically thin layer of carbon atoms. Second, neutral graphene is a
semimetal with zero density of states at  the Fermi energy and zero
gap, the two scales that shape the properties of metals and
semiconductors. Third,  the electronic structure close to the Fermi
energy has a conical shape with perfect electron-hole symmetry and
an internal  valley symmetry, isomorphic to that of two dimensional
massless Dirac fermions\cite{Semenoff84}.  As a result, most of the
standard lore on electronic and transport properties of low
dimensional semiconductors needs to be revisited
\cite{Katsnelson06,Beenakker06,Morozov06,Jarillo07}.

The presence of a band-gap in the electronic structure is crucial in the design
of low dimensional structures and in the achievement of large on-off resistance
ratios in field effect transistors (FETs).
 However,  even at very  low temperatures,
two dimensional graphene shows a rather low resistance at the charge
neutrality point\cite{Novoselov04,Geim05,Kim05,Jarillo07}  at which
the density of states at the Fermi energy is vanishingly small. A
number of physical mechanisms that result in a gap in graphene based
systems have been proposed: dimensional confinement into the so
called graphene ribbons
\cite{Fujita96,Wakabayashi99,Ezawa06,Brey06,PRBGuinea06}, interlayer
coupling \cite{exp-bilayer,AHM-DFT} and spin polarization
\cite{HMzz}. In this paper we focus on graphene ribbons, stripes of
graphene with finite width $W$. Ideal graphene ribbons with edges
along the crystallographic axis fall into two categories: zig-zag
and armchair \cite{Fujita96}. Only the former can present a gap
depending on their width $W$.  Recent transport experiments with
graphene ribbons \cite{Avouris} confirm that ribbons with $W=20nm$
present a thermally activated conductivity that indicates the
presence of  a gap. Most likely, imperfections on the edges of real
ribbons  make it hard to fabricate ribbons that are intrinsecally metallic.

In this paper we explore the relation between  gate voltage $V_G$
and injected density in  graphene-based field effect devices shown
in figure 1.  A graphene layer, of width $W$,  lies on top of an
insulating slab of thickness $d$ which lies above the metallic gate.
Application of  a gate voltage $V_G$ injects carriers in the
graphene layer accompanied by a corresponding change in the metal.
The main results of the paper are the following: {\em (i)} we show
that $V_G$ and the average two dimensional (2D) carrier density in
the system, $n_{2D}$ , satisfy the relation:
\begin{equation}
V_G= \frac{1}{C_{el}} en_{2D} +  V_{Q}(n_{2D})
\end{equation}
where the first term  arises from the classical electrostatic interaction and
the second term, $V_{Q}(n_{2D})$,
arises from quantum mechanical effects associated to the band
structure of graphene.  The classical contribution is dominant but
it depends only on the geometry  of the system. Although smaller,
the second term contains information specific to the electronic
structure of the system \cite{Mceuen06}. {\em (ii)} we find that in
2D graphene the quantum contribution scales like $V_{Q}\propto
\sqrt{n_{2D}}$, and from its measurement it would  be simple to
extract the linear slope of the graphene bands.  {\em (iii)} in the
case of ribbons we calculate the $V_G(n_{2D})$  curves for different
widths $W$ and we find the threshold gate above which semiconducting
ribbons become metallic. {\em (iv)} We argue that the ideal
conductance, $G$, in wide ribbons scales as $G=\sqrt{V_G}$

\begin{figure}
[t]
\includegraphics[width=3.0in]{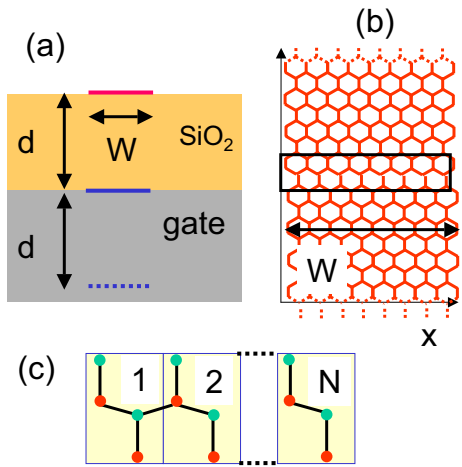}
\caption{ \label{fig1}(Color online). (a) Graphene based FET device: lateral
view. A graphene ribbon of width $W$ is separated from the metallic gate
by an insulating slab of
thickness $d$. The real charges of the metal are accumulated in the
metal-insulator interface underneath the ribbon.
Image charges lie further down the surface, at a
distance $2d$ from the graphene (b) Top view of an armchair graphene ribbon. The system is infinite along
the vertical direction ($y$ axis). (c) Detail of the super-unit cell defining
the periodic one dimensional armchair ribbon.   }
\end{figure}

The rest of this manuscript is organized as follows. In section II
we describe the Hartree formalism that we use to calculate the
electronic structure of gated graphene devices as well as the link
between gate voltage and chemical potential.  In section III we
present the simple analytical solution of the self-consistent
Hartree equation for planar graphene ($W=\infty$). We obtain an
expression for the quantum capacitance of graphene and discuss the
experimental implications. In section IV we present the numerical
solution of the self-consistent Hartree equation for  finite width
armchair ribbons, both semiconducting and metallic,  and we obtain
$n_{2D}$ and the conductance $G$ as a function of $W$,  and $V_G$.
In section V we discuss the implications and present our
conclusions.

\section{Theoretical framework }
\subsection{Hamiltonian}
We write the   Hamiltonian for electrons in graphene as the sum of two parts:
$H_0+U$. The first term describes the electronic structure of neutral graphene.
We approximate $H_0$ by the standard one-orbital tight-binding approximation
for both the 2D graphene layer \cite{Book} and graphene ribbons
\cite{Fujita96,Brey06,PRB2006US}.
The  model is completely defined by the positions of the atoms,
the first-neighbours hopping,
$t=-2.5 eV$, the lattice constant $a=2.42\AA$, and the
average density. The lattice of two dimensional graphene
is characterized by a crystal structure defined by the two unit vectors
$\vec{a}_{\pm}=\frac{a}{2}\left(\sqrt{3},\pm 1\right)$ with a two atom basis.
 We assume that the graphene
atoms lie in the $z=0$ plane. In the case of neutral graphene, the
average density is one electron per orbital that corresponds to a
Fermi energy $E_F=0$, since we take the on-site energy level equal
to zero. The armchair ribbons considered in this paper are generated by one
dimensional repetition of the super-cell of fig. 1b, along the $y$ axis, with
periodicity $\sqrt{3}a$. The supercell itself is constructed as the
repetition of along the direction $(1,0)$ of a block with 4 atoms
(figure 1c). If the supercell has $N$ blocks, the width of the
ribbon is given by $W=Na$ and the number of atoms in the unit cell
as  $4N$. As a rule of thumb, it is useful to write $W\simeq
(N/4)$nm.

The second  term $U$ describes the Coulomb interaction between the
extra charges in the system:
\begin{equation}
U= e \int  \delta \hat{n}(\vec{r}) \phi_{ext}(\vec{r}) d\vec{r}
+\frac{e^2}{2\epsilon}
 \int
\frac{\delta \hat{n}(\vec{r})\delta \hat{n}(\vec{r'})}{|\vec{r}-\vec{r}'|}
d\vec{r} d\vec{r'}
\end{equation}
where $\delta \hat{n}(\vec{r})$
is the operator describing the departure
of the local the electronic density from charge neutrality:
\begin{equation}
\delta\hat{n}(r)\simeq \sum_{I,\sigma} |\phi_I(r)|^2
\left(c^{\dagger}_{I,\sigma} c_{I,\sigma}-n_0\right)\, \, .
\label{deltan}
\end{equation}
Here the sum runs over all the atoms in the lattice and  $\phi_I(r)$
denotes the $\pi_z$ atomic orbital centered around the atom $I$ and
$c^{\dagger}_{I,\sigma}$ is the second quantization operator that
creates one electron with spin $\sigma$ in the orbital $\phi_I(r)$,
localized around atom $I$. In Eq.\ref{deltan} $n_0=1/2$ is the number of
electrons per site and per spin in neutral graphene.

The electrostatic  potential created by the extra charges in the metallic gate
is denoted by $\phi_{ext}(\vec{r})$. Considered as a whole,
 the graphene and the metal layer form a  neutral system. 
 Therefore, the extra carriers in one side are missing
in the other. Since the charges in the metallic gate move as to
cancel the electric field inside the metal, they depend on the
density distribution in the graphene electrode, which in turn
depends on $\phi_{ext}(\vec{r})$. This self-consistent problem is
solved using the image method: for a given charge density profile in
the graphene, $e\langle\delta \hat{n}(\vec{r})\rangle$ the potential
created by the corresponding extra charges in the metal is given by
a distribution of fictitious image charges inside the metal:
$-e\delta n_{im}(\vec{r})=-e\delta n(\vec{r}-2 \vec{d})$ with
$\vec{d}=(0,0,d)$.
\begin{equation}
V_{ext}(\vec{r})=-\frac{e}{\epsilon}
 \int \frac{\langle\delta \hat{n}(\vec{r'})\rangle}{|\vec{r}-\vec{r}'+2 \vec{d}|}
 d\vec{r'}\, \, \, .
\end{equation}
With this method the potential evaluated at the metal gate ($z=0$)
is exactly zero. Therefore, the potential difference between the
metal-insulator interface and the graphene layer is the potential
evaluated at the graphene layer.

\subsection{Hartree approximation}
The electronic repulsion is treated in the Hartree approximation.
Therefore, the electrons feel the
 electrostatic potentials
created by both the gate and themselves:
\begin{equation}
\hat{U}\simeq \hat{U}_{SC}=  e \int  \delta \hat{n}(\vec{r})
\left(V_{ext}(\vec{r})+V_{SC}(\vec{r}) \right)d\vec{r}
\label{hamil_2}
\end{equation}
The argument of the integral,  $V_T\equiv V_{ext}+V_{SC}$,
can be written as
\begin{equation}
V_T(\vec{r})=\frac{e}{\epsilon}
 \int \langle\delta n(\vec{r'})\rangle
 {\cal K}(|\vec{r}-\vec{r}'|,d)d\vec{r}'
  \end{equation}
where
\begin{equation}
{\cal K}(|\vec{r}-\vec{r}'|,d)=\frac{1}{|\vec{r}-\vec{r}'|}
 -\frac{1}{|\vec{r}-\vec{r}'+2\vec{d}|}
\end{equation}

In the Hartree approximation $H_0+U_{SC}$ is a one-body Hamiltonian
that can be represented  in the basis of localized
$\pi_z$ atomic orbitals. The $H_0$ part yields the standard single orbital
tight-binding Hamiltonian.
In contrast,  the  $U_{SC}$ introduces  site-dependent  shifts
in the diagonal matrix elements:
\begin{equation}
U_I=eV_I=\frac{1}{\epsilon} \sum_{J}  q_J v_{IJ}
\label{SCpot}
\end{equation}
where $q_J$ is the average excess charge in site $J$ and $v_{IJ}
=\int d\vec{r} d\vec{r}' |\phi_I(\vec{r})|^2 |\phi_J(\vec{r}')|^2
{\cal K}(|\vec{r}-\vec{r}'|,d)$, is the potential created by an unit
charge located at site  $J$ on site $I$.
At this point we adopt an approximation which permits to evaluate
$V_{IJ}$ without a detailed model of $\phi_I(\vec{r})$. Since the
atomic orbitals are highly localized, for $|R_I-R_J|>a$ we can
approximate  $v_{IJ}$ by the first term in the multipolar expansion:
$v_{IJ} \simeq  \frac{1}{|\vec{R}_I-\vec{R}_J|}
-\frac{1}{|\vec{R}_I-\vec{R}_J-2 \vec{d}|}$
We have verified that the potential in a given site is quite independent of the
approximation adopted to evaluate the $I=J$ contribution, for which the
monopolar approximation fails. Without sacrificing
accuracy, we adopt the simplest strategy of removing that term from the sum.

The Sch\"rodinger equation reads
\begin{equation}
(H_0+U_{SC})|\alpha\rangle=\epsilon_{\alpha}|\alpha\rangle
\label{scho}
\end{equation}
where the eigenvectors $|\alpha\rangle$ are  linear combinations of
atomic orbitals: $|\alpha\rangle=\sum_I {\cal C}_{I,\alpha}
|\phi_I,\sigma\rangle $, and the coefficients are independent of the
spin because we  only consider non-magnetic solutions. In turn, the
distribution of excess charges depends on the eigenvalues and
eigenvectors through the equation
\begin{equation}
q_J=e\left[\sum_{\alpha,\sigma}
\left(|{\cal C}_{I,\alpha}|^2 f(\epsilon_{\alpha})\right)-1\right]
\label{SCcharge}
\end{equation}
where $f(\epsilon_{\alpha})$ is the Fermi function.

Equations (\ref{SCpot}), (\ref{scho}) and (\ref{SCcharge})
need to be solved iteratively.
In each step of the iterative procedure we have an {\em input}
distribution of extra charge, $q_I$  which results in a
 potential (eq. (\ref{SCpot})), which defines a
Hamiltonian whose eigenstates (eq. (\ref{scho})) result in an {\em output}
distribution of
extra charge (eq. (\ref{SCcharge}). The converged solution is such that the input and output
distribution of charges are the same. We refer to that distribution and the
corresponding Hamiltonian as the self-consistent solution.

\subsection{Gate voltage,  chemical potential and quantum capacitance}

In real field effect devices,   application of a gate voltage $V_G$
results in a change of carrier density in the active layer. In our
theoretical framework, we  define $eV_G$  as the  the chemical
potential difference between the metallic backgate and the graphene
ribbon,  necessary to accommodate extra carriers (either holes or
electrons) in graphene and remove them from the metallic backgate:
\begin{equation}
eV_G(n_{2D})= \mu_{graphene}-\mu_{metal}
\end{equation}
where $n_{2D}$ is the average two-dimensional density  
of {\em extra} carriers. 
In general, there are two contributions to
the  dependence of the  chemical potential on $n_{2D}$. One one
side, the presence of the electrostatic potential created by extra
carriers present both in graphene and the metallic backgate shifts
and modifies the bands. We refer to the new bands as the Hartree
bands. On the other, addition of new carriers involves an additional
shift of the chemical potential with respect to the Hartree bands.
Whereas the first contribution has a classical origin, the second is
a consequence of the Pauli principle and we refer to it as quantum
contribution to the capacitance.  We neglect a   third type of
contribution , arising from the
modification of the bands due to density dependent exchange and correlation
contributions\cite{Mceuen06}.

The extra carriers in the metallic gate are concentrated on the
surface so that the  {\em bulk} chemical potential  and the {\em bulk} energy
bands are shifted by the same amount: there is no quantum
contribution to the capacitance coming from the bulk metal.
Things are different in
the graphene layer. In this case,  it makes no sense to talk about
bulk and surface as different objects. In graphene, the  chemical
potential has to shift relative to the modified bands in order to
accommodate extra carriers, resulting in a non-zero quantum
contribution to the capacitance. Without loss of generality we
take the metallic side as the reference for
the electrostatic contribution to the chemical potential. This
permits to write
\begin{equation}
eV_G=\mu_{graphene}(n_{2D})-\mu_{graphene}(0)\equiv\delta\mu
\end{equation}
From an operational point of view, we consider the system to  be at
zero temperature, and we  take the chemical potential of graphene at
a given density as the lowest unoccupied eigenstate of the
self-consistent Hamiltonian. Identical results are obtained from the
equation
\begin{equation}
n_{2D} = \int_{-\infty}^{\mu} \rho_{SC}(E) dE
\label{DOS-SELF}
\end{equation}
that relates the chemical potential, the extra density via the
density of states of the self-consistent
Hamiltonian, $ \rho_{SC}(E)$.


\section{Classical and quantum capacitance of planar graphene}
In the case of 2D graphene all the atoms of the lattice are equivalent and
therefore the extra charge $q_I$ and the effective potential $U_I$ are
independent of the location. As a matter of fact, for an
 average extra density  $n_{2D}$, the electrostatic potential
 is $U_I= \frac{4\pi e^2}{\epsilon}  n_{2D}d$.  Therefore,
 the Hartree bands of charged
 graphene  are equal to the bare bands plus a
 rigid shift:
 \begin{equation}
 E_{\nu}(\vec{k})=  \epsilon^{0}_{\nu}(\vec{k})
 + \frac{4 \pi e^2}{\epsilon}  n_{2D} d
 \end{equation}
In figure 2 (left panels) we plot both the bare bands
$\epsilon^{0}_{\nu}(\vec{k})$ and the Hartree bands
$E_{\nu}(\vec{k})$ for $d=300$nm and a $n_{2D}=10^{12}$cm$^{-2}$. We
see that Hartree bands are shifted upwards. Accordingly, the shift
in the chemical potential needed to accommodate this extra charge
will have to  account for the electrostatic shift of the bands.
On top of that, the chemical potential needs to move away from the Dirac point,
which is also shifted electrostatically. In order to calculate this second
contribution, we restrict ourselves to the energy
region around $E_F=0$ in which the density of states is linear (linear bands in
2D). This is a very good approximation since
we consider $n_{2D}<< n^0\simeq 3.9$ 10$^{15}cm^{-2}$, the equilibrium
electronic density in graphene.
 Therefore we write the density of states of the self-consistent Hamiltonian:
 \begin{equation}
\rho_{SC}(E)=\rho_0\left(E-\frac{4 \pi e^2}{\epsilon}n_{2D} d\right)
\end{equation}
with:
 $\rho_{0}(E)= \frac{8}{3\pi t^2a^2} |E|$

After some simple steps we write the equation relating the change in chemical
potential $\delta \mu$
\begin{equation}
n_{2D}=
\int_{0}^{\delta \mu} \rho_{SC}(E) dE
\end{equation}
which yields
\begin{equation}
n_{2D} =\frac{8}{3\pi} \frac{1}{2 t^2a^2}
\left|\delta \mu-\frac{4 \pi e^2}{\epsilon}n_{2D} d\right|^2
\end{equation}
From here we obtain one of the important results of this work:
\begin{equation}
eV_G =\frac{4 \pi e^2}{\epsilon}d n_{2D} +
|t| \sqrt{\frac{6\pi n_{2D} a^2}{8}}\equiv
\frac{e n_{2D}}{C_{el}}+eV_Q
\label{Vgate2D}
\end{equation}

\begin{figure}
[t]
\includegraphics[width=3.0in]{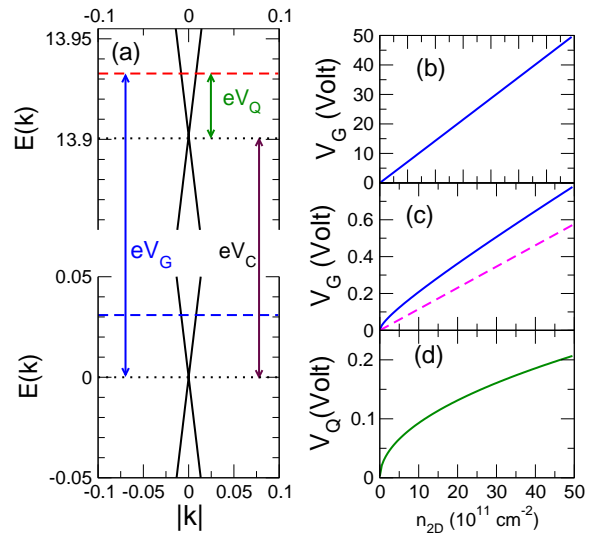}
\caption{ \label{fig2}(Color online).  {\bf (a)}:
Bare (below) and Hartree (above)
bands, close to the Dirac point,  of two
dimensional graphene with $n_{2D}=10^{12}$cm$^{-2}$.
 The vertical lines represent
 the chemical potential of the charged system, $eV_G$, which is
the sum of two contributions, $eV_C$, the rigid shift of the bare
bands to yield the Hartree bands, and $eV_Q$, the shift of the
chemical potential with respect to the Hartree bands, the quantum
contribution to the gate. Dotted horizontal lines represent the bare
and charged Dirac points and dashed horizontal lines the bare and
charged chemical potential. {\bf (b)} $V_G=V_C+V_Q$ versus density
for 2D graphene in a FET with $d=$300 nm and $\epsilon=3.9$.
{\bf (c)}  $V_G=V_C+V_Q$ (solid line) and $V_C$ (dashed line)
versus density for 2D graphene in a FET with $d=$30 nm and $\epsilon=47$ (see
for instance \onlinecite{hf})
  {\bf (d)} Quantum contribution to the gate, $V_Q$ versus $n_{2D}$.
  }
\end{figure}

Thus, we write the gate voltage as the sum of two terms.
The first
is the standard  electrostatic contribution whereas the second
is related to the density of states of graphene. We refer to the second, $V_Q$,
as the the quantum contribution to the gate voltage (or inverse capacitance).
In most of  devices so far\cite{Novoselov04,Geim05,Kim05,Jarillo07}
 $d\simeq$300 $nm$ ,which makes the electrostatic
contribution much larger than the quantum contribution.
In figure 2b we plot $V_G$ vs $n_{2D}$,
according to equation (\ref{Vgate2D}). In spite of the last term of eq.
(\ref{Vgate2D}), the
curve looks like a straight line. As  in  refs. (\onlinecite{Geim05,Kim05})
we have done a linear fit of fig. 2b  obtaining $n_{2D}=\alpha
V_G$ with $\alpha=0.718 $ 10$^{11} V^{-1}$ $cm^{-2}$, in very good
agreement with the experiments.
In order to reduce the electrostatic contribution as much as possible, in figure
2c  we
consider a thinner dielectric ($d=$30 nm) with a much larger dielectric constant
$\epsilon=47$ \cite{hf}. We see how in this case
the gate voltage (solid line) is signifcantly different from 
the electrostatic contribution (dashed line).  
In figure 2d we plot the quantum contribution, which is independent of
$\epsilon$ and $d$, alone.
Although much smaller than the classical term, it can be larger than
0.2 Volt and it should be possible to measure it. Therefore, the independent
measurement of $n_{2D}$, via classical Hall effect, and $V_G$ could provide
a direct measurement
of the slope of the  bands in the linear region if the experimental results are
fitted using eq. (\ref{Vgate2D}).


\section{Self-consistent electronic structure of arm-chair ribbons}

\subsection{Bands and charge profiles}
We now consider finite width armchair ribbons. They are different
from two dimensional graphene both because their electronic
structure and their electrical capacitance. Armchair ribbons are
metallic (figure 3a,c)  when the  width of the sample has the form
$W$=$(3M+1)a_0$, with $M$ an integer, and insulating otherwise
(figure 3b,d). In both case the electronic structure is different of
the semimetallic behaviour of 2D graphene. On the other side, the
presence of edges breaks translational invariance along the
direction perpendicular to the ribbon axis. The edges provide a
natural surface to accommodate the extra electrons. Therefore,  the
local potential and the extra charge  profile are expected to vary
as a function of the distance to the ribbon edge, keeping their
translational invariance along the ribbon axis. This makes it harder
to separate classical and quantum contributions to the capacitance
in simple terms.

\begin{figure}
[b]
\includegraphics[width=3.0in]{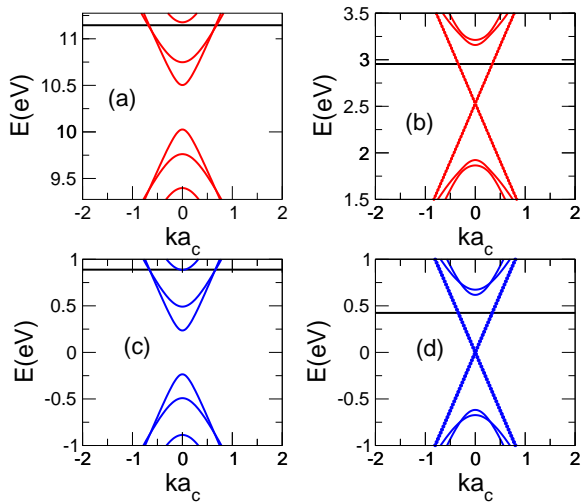}
\caption{ \label{fig2}(Color online). Energy bands for $N=9$ (left
panels) and $N=10$  (right panels). The Hartree (bare) bands are
shown above (below). The  corresponding densities are $n_{2D}= 7.6
10^{12}cm^{-2}$ (N=9) and $n_{2D}= 1.7 10^{12}cm^{-2}$ (N=10).
Notice that the neutral and Hartree bands are shifted with respect
to each other. The horizontal lines indicates the Fermi energy.}
\end{figure}

In figure 3  we show both the self-consistent  (upper panels)  and the
neutral (lower panels) bands for
electrically doped ribbons in two cases,   semiconducting and a
metallic ($N=9$ and $N=10$ respectively). Notice that charge neutral
bands have
electron-hole symmetry, like the semiconducting carbon nanotubes \cite{Jarillo1}.
 In those figures we plot
the self-consistent chemical potential in the upper panels and the
naive Fermi  energy  obtained upon integration of the neutral density of states.
As in the case of 2D graphene, the electrical injection of extra carrier results
in a large shift of the bands, due to the electrostatic interactions.
In contrast to the 2D case, the inhomogeneity of the electronic density and
the electrostatic potential result in a moderate change of the
the shape of the bands. Notice for instance that for the $N=9$ ribbon
 the chemical potential
intersects 2 self-consistent bands whereas the naive Fermi energy intersects 3
neutral bands.

As in the two dimensional case, the self-consistent chemical
potential can be though as the sum of  a  large electrostatic shift
and a smaller quantum shift with respect to the self-consistent
bands, necessary to to accommodate the extra  carriers. However, the
electrostatic contribution to the gate depends not only on the
average to density $n_{2D}$, like in the 2D case, but also on the
detailed profile $q_I$. As a result, the separation of the chemical
potential in two contributions, electrostatic and quantum
mechanical, does not yield a simple procedure to obtain information
of the bands of gated graphene ribbons, in contrast to the simpler
2D case.

\begin{figure}
[hbt]
\includegraphics[width=3.0in]{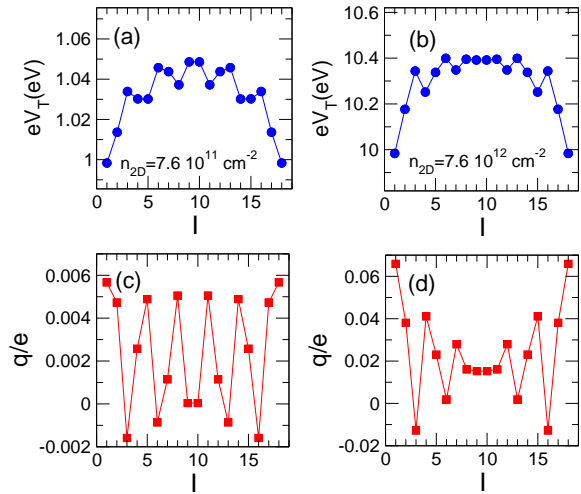}
\caption{ \label{fig4}(Color online).
 Potential ((a),(b)) and density ((c),(d))profiles, as function of
 the position across the ribbons,
 for $N=9$ ribbon with $n_{2D}=7.6$
10$^{11}cm^{-2}$ (a,c) amd $n_{2D}=7.6$ 10$^{12}cm^{-2}$ (b,d)
 The higher
density results correspond to the bands shown in figure 3a. }
\end{figure}

In figure 4 we show  the self-consistent density profile $q_I$ and
the self-consistent electric potential $U_I$ for a the
semiconducting ribbon $N=9$ with two different $n_{2D}$. We see how
the density of carriers is larger in the edges than in the middle of
the ribbon, as expected in a conducting system. As a result, the
electrostatic potential has a inverted U shape. Superposed with
these overall trends, both $q_I$ and $U_I$ feature oscillations,
arising from quantum mechanical effects. Expectedly, the average
$U_I$ is much larger in the high density than in the low density. In
the high density case it is apparent that the potential is flat in
the inner part of the ribbon, very much like in a metal. Very
similar trends are obtained for ribbons with different widths.

\subsection{Density vs $V_G$ }
In figure 5a we plot the gate versus the average 2D extra density in
graphene ribbons with for different widths $N$, indicated  in the
figure. This can be a valuable information to estimate the carrier
density in gated ribbons since Hall measurements in narrow ribbons
might be difficult to perform. The common feature in all the curves
is the linear relation between $V_G$ and $n_{2D}$ that reflects the
dominance of the classical electrostatic contribution over quantum
effects, exactly like  in the 2D case. This is also apparent from
figure 3, where the shift of the bands is much larger than the Fermi
energy with respect to the bottom of the conduction bands. In
addition, we see that some of the curves do not intersect at $V_G=0$
for $n_{2D}=0$ in the case of semiconducting ribbons. This is
clearly the case of $N=9$ and $N=20$. In contrast, the curve $N=10$
extrapolates to zero. We denote the threshold gate potential as
$V_{th}$. and we notice that it  corresponds to the the change in
chemical potential of the ribbon when a single electron is added.
Since the chemical potential for a semiconductor system lies in the
middle of the gap and the chemical potential when a single electron
is added is the lowest energy state of the conduction band, we must
have  $eV_{th}=\frac{E_g}{2}$ where $E_g$ is the gap of the
semiconducting ribbon. We fit all the curves in figure 5a according
to:
\begin{equation}
n_{2D}(N)= \alpha(N)\left(V_G -V_{th}(N)\right)
\end{equation}
In figures 5b and 5c we plot $\alpha$ and $V_{th}$ as a function of the ribbon
width, $N$. The wider ribbon considered has $N=60$ which corresponds to
$W=14.5$nm.
 In figure 5b we see how
$\alpha$ rapidly decreases towards the 2D value $\alpha= \frac{e^2
d}{\epsilon}$=0.7 $10^{11} cm^{-2} V^{-1}$ as $W$ increases.
We have verified that $\alpha$ scales inversely proportional to the width $N$.
Interestingly,  $\alpha(N)$ approaches to the two
dimensional case very quick, even if $W<<d$. The evolution of
$V_{th}$ as a function of $N$ reflects two facts: on one side,  2
out of 3 ribbons are semiconducting. On the other, the gap decreases
as $W^{-1}$.  The evolution of the gap as a function of $W$ in the
one-orbital tight-binding calculation similar, but not identical, to
that  obtained with density functional calculations \cite{HMzz}.
Notice that ribbons of $W=20nm$ ($N\simeq 80$) still present a gap
\cite{Avouris} and, according to figure 5b, their capacitance is
almost that of two dimensional graphene.

\begin{figure}
[hbt]
\includegraphics[width=3.0in]{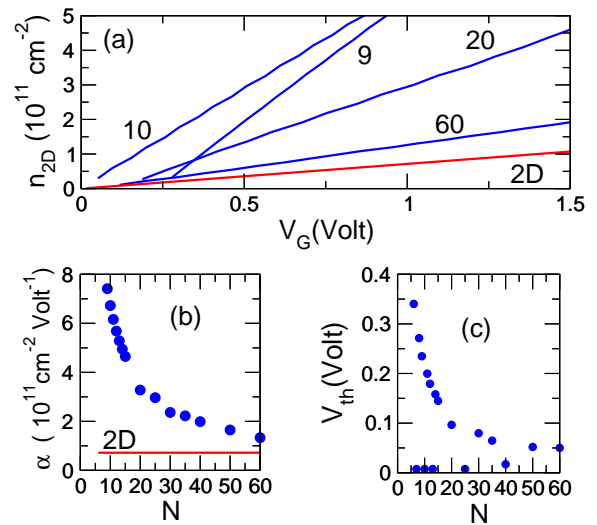}
\caption{ \label{fig4}(Color online). {\bf a} Gate voltage versus
average density for armchair ribbons of different widths. {\bf b}
Inverse capacitance, $\alpha$, as a function of width (measured in
terms of $N$).  {\bf c} Threshold voltage as a function of the
ribbon width $N$. }
\end{figure}

\subsection{Ideal Conductance vs  $V_G$ }

We now study the number of occupied bands ${\cal N}$ in a given ribbon as a function
of $V_G$. This is related to ideal  conductance
a function of $V_G$ through
 the Landauer formula for perfectly transmitting ribbons:
\begin{equation}
G=\frac{2e^2}{h} {\cal N}
\end{equation}
This is equivalent to neglect the effect of disorder in the system.
Therefore, the  value of conductance so obtained can be considered an upper
limit for the real conductance in the system \cite{PRB2006US}. In an ideal
armchair ribbon the number of chanels ${\cal N}$ increases in one by one as Fermi energy
with respect to the Dirac point is
increased. Our approach
permits to obtain both ${\cal N}$ and the quantum shift of the chemical
potential as a function of the gate voltage and to plot ${\cal N}$ as a function
of $V_G$.

It must be stressed that in the 2D case this disorder free
model\cite{Beenakker06}   accounts for  the experimental
results\cite{Geim05}. This might indicate that either disorder is
not present in the samples or, more likely,  because of the
suppression of the backscattering it does not affect significantly
the transport properties of graphene\cite{Beenakker06}. Transport on
2D graphene is being extensively studied by a number of groups.  The
effect of imperfections on the transmission of otherwise ideal
graphene ribbons has also been studied by a number of
groups\cite{PRB2006US}.

\begin{figure}
[hbt]
\includegraphics[width=3.0in]{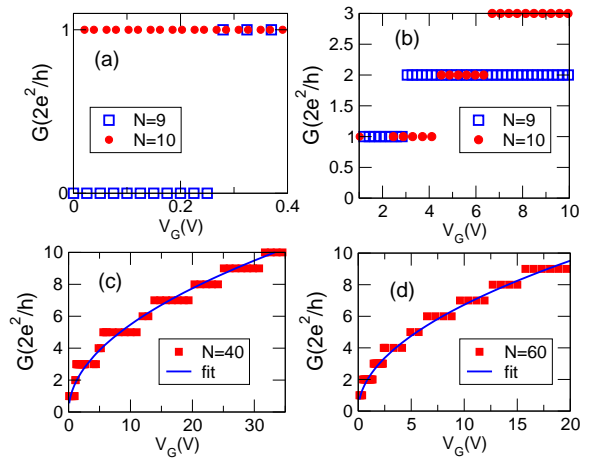}
\caption{ \label{fig4}(Color online). Conductance as a function of
$V_G$ for ribbons with $N=9$ and $N=10$ (a) and (b), $N=40$ (c) and
$N=60$ (d) (see text)}
\end{figure}

In figure 6 we plot $G(V_G)$ for various ribbons. In fig. 6a we plot the
conductance $G(V_G)$ for two narrow ribbons ($N=9$ and $N=10$) in the small
gate regime. It is apparent that the $N=9$ ribbon only conducts above a
threshold gate  whereas the $N=10$ ribbon is conducts even for $V_G=0$.
Therefore, semiconducting ribbons  can be electrically tuned from insulating to
conducting behaviour  with a gate voltage. This is also different from the case
of 2D graphene which has a rather low resistance in the charge neutrality point.
In figure 6b we show the conductance for the same ribbons at higher $V_G$.
The different sizes of the plateus for $N=9$ and $N=10$  reflect the different
structure of the bands, as seen in figure 3. In figures 6c and 6d we show the
conductance for wider ribbons ($N=40$ and $N=60$ respectively).
Although  the shape of
the curves is superficially similar for the all these ribbons,
the $V_G$ necessary
to have a fixed number of bands at the Fermi energy is a decreasing function of
$W$. This is a consequence of quantum confinement: the smaller the ribbon the
larger the sub-band level spacing, $\Delta E_n=E_n(k=0)-E_{n-1}(k=0)$. A shift
of the Fermi energy, relative to the self-consistent bands,  by an amount $\Delta
E_n$, so that a new band is available for transport,  implies also an
electrostatic overhead, due to the change in density, which accounts form most
of the $eV_G$, as we have discussed above.

In average, the steps in the stepwise curve ${\cal N}(V_G)$ increase size as $V_G$
increases. This trend is more apparent in wider ribbons.
In this sense  it can be said that  ${\cal N}$ is sublinear in
$V_G$. In the
limit of very wide ribbons, for which the density of states is almost 2D,
 we can derive a qualitative relation between number
of channels and $V_G$.  Since the number bands per unit energy is roughly
constant,  we have that ${\cal N}$ scales linearly
with the Fermi energy with respect to the bottom of the bands,
${\cal N}\propto V_Q$. On the other side, we know that the density of carriers
scales linearly with $V_G$ (figs. 2 and 5) and, using the 2D density of states
we know that  density of carriers scales with
$V_G\propto n\propto V_Q^2$. Therefore, we conclude
\begin{equation}
{\cal N}\propto \sqrt{V_G}
\label{GsqV}
\end{equation}
In figures 6c and 6d we have fitted the numerical data to the curve $G=a
V_G^b$, where $a$ and $b$ are fitting parameters.
We have obtained exponents $b=0.45$ and $b=0.47$ for $N=40$ and $N=60$
respectively, in agreement with the qualitative discussion above.  In figure 6c
and 6d we the solid line is the best fit to the equation $G=a
\sqrt{V_G}$. Equation
(\ref{GsqV})  is a re-statement of the  relation between $V_Q$ and $n_{2D}$ of
eq. (\ref{Vgate2D}). In both cases the exponent $1/2$ comes from the linear
relation between energy and momentum of the two dimensional carriers on one
side, and the quadratic relation between $n_{2D}$ and Fermi momentum on the
other. Therefore, eq. (\ref{Vgate2D}) and (\ref{GsqV}) are specific predictions
related to the peculiar Dirac-like spectrum of electrons in graphene.

\section{Discussion and Conclusions}
In this work we study how the electronic structure and the density of carriers
of graphene based field
effect transistors evolve as a gate voltage $V_G$ is applied.
An important notion is the identification of $eV_G$  as the change in chemical
potential in the graphene layer necessary to accomodate the density of extra
carrers $n_{2D}$.  The change in chemical potential is the sum of a
electrostatic contribution, independent of the details of the electronic
structure of graphene and a quantum contribution that depends on the details of
the  graphene band structure.  In the case of 2D graphene, we have
obtained a particularly simple equation (\ref{Vgate2D})
relating these quantities. We
find that the quantum contribution is sufficiently small as to go unnoticed in
the experiments done so far \cite{Novoselov04,Kim05,Geim05},
but sufficiently big as to be measured.
Even more important, we claim that an independent measurement of $n_{2D}$ and
$V_G$  would provide a direct
measurement of the slope of the linear energy bands,
the so called graphene "speed of light".

We have considered idealized armchair ribbons which, at the charge neutrality
point, can be either semiconducting or metallic. Application of a $eV_G$ equal
to half the band-gap turns semiconducting ribbons into conductors. This affords
a on/off ratio much larger than that of 2D graphene.
In the case of  graphene ribbons the classical contribution to capacitance
depends on the charge density profile $q_I$ in the ribbon (fig. 3),
which can not be measured easily, and we can not provide a simple experimental
procedure to extract information about the electronic structure of the ribbons
out of the $V_G(n_{2D})$ curves.  We find that density of extra carriers is
higher in the edges of the ribbons than in the middle.

Future work should address the issue of the stability of the results obtained
in this paper with respect to disorder. In particular, we expect that disorder
will turn metallic ribbons into  semiconducting at $V_G=0$. In figure 3 it is
apparent that the metallic character of the ribbon with $N=10$  comes from the
states lying in a narrow window of width $\Delta k_x$ around $k_x=0$ in
momentum states.  These states would dissappear if size of the ribbon along the
$x$ axis is smaller than $\frac{1}{\Delta k_x}$.

In conclusion, we have studied the electronic structure of gated graphene and
graphene ribbons in the Hartree approximation. This permits to include the
Coulomb repulsion between the extra carriers and their coupling to the external
gate in a selfconsistent manner.
We find (eq. (\ref{Vgate2D}) a small departure from
the classical linear result
$V_G\propto n_{2D}$ which can provide a simple method to measure the slope of
the graphene bands.  In the case of semiconducting armchair ribbons we have
obtained the inhomogeneous distribution of charge carriers $q_I$ and potential
$U_I$ along the section of the ribbon. We also  find that a finite  $V_G$ equal
to half their band-gap is necessary to make them conduct.

We acknowledge useful discussions with H. Fertig, P. Jarillo-Herrero, F.
Guinea, C. Untiedt, D. Jacob and F. Mu\~noz-Rojas. This work has
been financially supported by MEC-Spain (Grants FIS200402356,
MAT2005-07369-C03-01, MAT2006-03471 and the Ramon y Cajal
Program), and by Generalitat Valenciana (GV05-152). 
This work has been partly funded by FEDER
funds.

\widetext
\end{document}